# Shot-to-shot spectrally-resolved characterization of continuous-wave-triggered supercontinuum near 1μm

**Zhibo Ren, Yi Qiu, Kenneth K. Y. Wong, Kevin Tsia***
*Department of Electrical and Electronic Engineering, The University of Hong Kong, Pokfulam Road, Hong Kong, P. R. of China*
*Corresponding author: tsia@hku.hk*

**Abstract:** We demonstrate shot-to-shot (at a rate of 20 MHz), broadband (>200 nm), spectrally-resolved (spectral resolution of 0.2 nm) statistical characterization of continuous-wave-triggered supercontinuum in the 1 μm regime, enabled by optical time stretch.

The ultrabroad bandwidth is the hallmark of the supercontinuum (SC) light which has long been identified to benefit a diverse range of applications, such as frequency metrology and biomedical diagnostics. Another SC characteristic, the noise property, has also been progressively drawing attention – mainly driven by the growing demand for practical low-noise SC. The well-established theoretical framework of SC generation has been extensively employed to gain more insight on the stochastic nature of SC. However, the mechanism of such random fluctuation is known to be complex and sensitive to various intertwined factors, such that the characteristics of the input pump pulse as well as, if any, the weak controllable perturbation, such as a continuous-wave (CW) trigger [1,2]. It is thus of great importance to perform direct measurements of the *real-time spectrally-resolved dynamics* of SC generation process in the actual experimental settings. It is yet not a straightforward task due to the lack of real-time single-shot spectrometer. Previously applied in ultrafast spectroscopy and imaging [3], optical time-stretch (OTS) technique was recently found to be a viable approach to experimentally investigate the shot-to-shot spectral fluctuation and correlation of SC in real time [4]. OTS maps the spectrum of each SC pulse into a temporal waveform by group velocity dispersion (GVD). The waveform in essence is the replica of the spectrum in time, allowing spectral acquisition by a high-speed photodetector at a speed not achievable by the conventional spectrometers. In this paper, we further validate and expand its utility to experimentally characterize the real-time intensity stability and the spectral correlation of CW-triggered SC in the 1μm regime with a measurement bandwidth as wide as ~200 nm. The experimental results are found to be in good agreement with the numerical simulation based on generalized nonlinear Schrödinger equation (GNSLE). We also show that real-time, shot-to-shot, broadband SC characterization by OTS is feasible not only limited in the telecommunication band [5], but also in the 1μm range – an important window for many biomedical imaging and spectroscopy applications. The present work also shows that OTS represents a handy approach to experimentally identify the favorable active-control scenarios (by using a CW trigger) for generating SC with the desired performance (e.g. bandwidth, spectral stability and spectral intensity correlation).

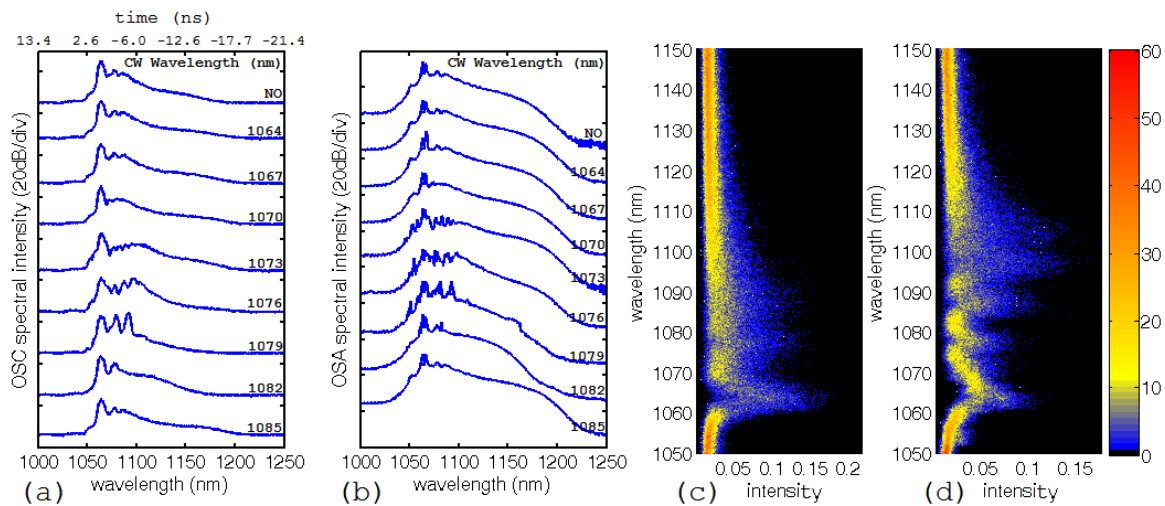

Fig.1. (a) Averaged OTS spectra. (b) Spectrum measured by the OSA. (c) – (d) Untriggered SC and 1076nm CW-triggered SC histograms across 1050 nm – 1150 nm, respectively. Time axis is also shown in (a) to show the wavelength-to-time. Histograms are obtained from 399 time-stretched SC pulses. The color bar denotes the number of events in each linear wavelength and intensity bins.

In our experiment, a 20 m-long photonic crystal fiber (PCF) is pumped in the anomalous dispersion regime by a picosecond mode-locked laser pulse centered at 1064 nm (FWHM: ~7 ps, peak power: ~80 W). A weak CW trigger (~320 times smaller than pump power) from a wavelength tunable CW laser (1064 - 1085 nm) is launched into the PCF together with the pump. The output SC pulses are subsequently time-stretched by a single-mode fiber (GVD: ~0.18 ns/nm, corresponding to a spectral resolution of ~0.2 nm [3]) and captured by a real-time oscilloscope (16 GHz, 80 GSa/s). Fig. 1(a) shows the averaged OTS spectra (using 399 consecutive single-shot spectra captured at a rate of 20MHz, governed by the repetition rate of the laser) at each CW trigger wavelength. They are in good agreement with the spectra acquired by an optical spectrum analyzer (OSA) at a scan rate of ~2 Hz (Fig. 1(b)). Both the SC bandwidth and the spectral features are strongly influenced by varying the trigger wavelength. In particular, it is well-known that the stochastic nature of soliton fission tends to wash out the spectral variation in the averaged SC spectrum (see untriggered SC in Fig.1 (a) – (b)). This is in clear contrast to the CW-triggered SC spectra in which the spectral undulation around 1064–1100 nm, signifying the higher-order soliton fission with higher temporal stability, becomes more pronounced when the CW-trigger wavelength is around 1076 – 1079 nm (see Fig.1 (a) – (b)). We also stress that the OTS allows us to access of both the statistical and spectral information of the SC in real-time. This can be evident from the *spectrally-resolved* intensity histograms shown in Fig. 1(c)-(d). Clearly, we can observe the key effects of the CW-triggered SC: (1) spectral enhancement with additional undulation feature, and (2) modest improvement in temporal stability on the longer wavelength side of the SC. This is consistent with the recent findings based on numerical simulation that soliton fission is initiated in a more deterministic fashion in the presence of the weak CW-trigger [1]. By evaluating the intensity spectral correlation of the time-stretched CW-triggered SC spectra (Fig.2(b), CW trigger wavelength of 1076 nm), we also observe higher intensity correlation (i.e. either higher positive or negative correlation) near 1064 – 1100 nm, where the soliton-like features are clearly apparent. In the untriggered SC case (Fig. 2(a)), the spectral correlation drops rapidly away from the pump, indicating the spectral decoherence due to the noise-seeded modulation instability (MI), which leads to stochastic onset of soliton fission, and thus SC.

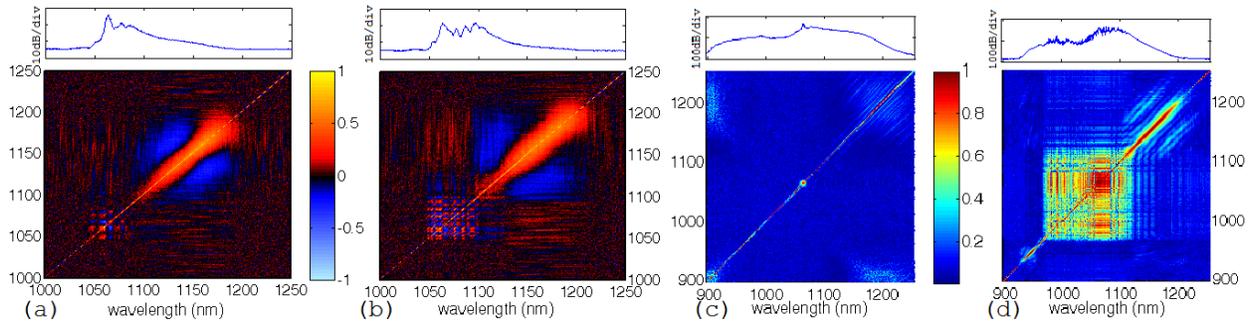

Fig.2. Spectral intensity correlation maps of time-stretched (a) untriggered SC, (b) 1076 nm CW-triggered SC. The color bar denotes the correlation values, between -1 to 1. The averaged OTS spectra also shown on top of the maps for clarity; Simulated CSD function of (c) untriggered SC, (d) 1076 nm CW-triggered SC. Color bar represents the coherence, ranging from 0 (incoherent) to 1 (coherent).

It is in accord with our coherence analysis based on the two-frequency cross spectral density (CSD) function from GNLSE simulation. By introducing a CW trigger at 1076 nm, the CSD reveals higher coherence within a square region in the range of 975 nm – 1125 nm (Fig.2 (d)). Without the CW-trigger, only the diagonal line with high coherence is visible in the CSD (Fig.2(c)). The square is classified as the nearly coherent components, whereas the diagonal line corresponds to the quasi-stationary components of the SC [6]. Hence, it is clear that CW-triggering at proper wavelength (e.g. 1076 nm in the MI gain spectrum) improves the SC coherence.

In summary, we demonstrate real-time shot-to-shot broadband (~200 nm) experimental characterization of CW-triggered SC, by OTS in 1μm regime – exploiting a wider range of spectral window to perform comprehensive investigation of SC, particularly its intrinsic shot-to-shot noise properties. We note that OTS enables us to assess the spectrally-resolved statistics of the SC under the influenced of CW triggering. It thus could represent an effective experimental tool to search for optimal conditions for actively manipulating SC properties.

The work in this paper is partially supported by grants from the RGC of HKSAR, China: HKU 717510E, HKU 717911E, HKU 720112E, HKU 7172/12E, and University Development Fund of HKU.